\documentclass{article}

\usepackage{PRIMEarxiv}

\usepackage[utf8]{inputenc} 
\usepackage[T1]{fontenc}    
\usepackage{hyperref}       
\usepackage{url}            
\usepackage{booktabs}       
\usepackage{amsfonts}       
\usepackage{nicefrac}       
\usepackage{microtype}      
\usepackage{fancyhdr}       
\usepackage{graphicx}       
\graphicspath{{media/}}     
\usepackage{float}
\usepackage[ruled,vlined]{algorithm2e}
\usepackage[symbol]{footmisc}

\title{Real-Time Construction Algorithm of Co-Occurrence Network Based on Inverted Index
}

\author{
  Jiahao Cheng \\
  Department of Management Science and Information Management \\
  Wuhan University of Technology \\
  Wuhan, Hubei, China\\
  \texttt{jaydencheng@whut.edu.cn} \\
}

\begin{document}
\maketitle

\begin{abstract}
Co-occurrence networks are an important method in the field of natural language processing and text mining for discovering semantic relationships within texts. However, the traditional traversal algorithm for constructing co-occurrence networks has high time complexity and space complexity when dealing with large-scale text data. In this paper, we propose an optimized algorithm based on inverted indexing and breadth-first search to improve the efficiency of co-occurrence network construction and reduce memory consumption. Firstly, the traditional traversal algorithm is analyzed, and its performance issues in constructing co-occurrence networks are identified. Then, the detailed implementation process of the optimized algorithm is presented. Subsequently, the CSL large-scale Chinese scientific literature dataset is used for experimental validation, comparing the performance of the traditional traversal algorithm and the optimized algorithm in terms of running time and memory usage. Finally, using non-parametric test methods, the optimized algorithm is proven to have significantly better performance than the traditional traversal algorithm. The research in this paper provides an effective method for the rapid construction of co-occurrence networks, contributing to the further development of the Information Organization fields.
\end{abstract}


\section{Introduction}
With the development of the big data era, fields such as information organization, knowledge graphs, and library science have gained increasing attention. Co-occurrence networks, as effective tools for revealing potential associations between words in text \cite{a1, a2, a3}, play a crucial role in various applications, including text mining, sentiment analysis, knowledge representation, and recommendation systems. Co-occurrence networks are graph structures with words as nodes and co-occurrence relationships as edges, enabling the discovery of latent semantic associations and facilitating efficient information organization and management \cite{a4}.

However, the current methods for constructing co-occurrence networks suffer from high time and space complexity, making it challenging to handle the processing demands of large-scale data and real-time applications. This problem becomes more pronounced in the era of big data. From an engineering perspective, this limitation hinders the integration of co-occurrence network analysis into real-time systems, such as the Web. In the context of modern information retrieval, knowledge discovery, and software cloudification, real-time and dynamic characteristics are crucial, highlighting the limitations of existing methods.

This paper first introduces the concept and main applications of co-occurrence networks. It then introduces the inverted index data structure and discusses the principles and limitations of current algorithms in conjunction with application scenarios. To address these limitations and optimize performance, this paper proposes a novel high-performance real-time co-occurrence network construction algorithm by incorporating inverted index and breadth-first search. Experimental testing and comparative analysis using an academic paper metadata dataset are conducted to demonstrate the superiority of the proposed algorithm.

\section{ Related Concepts and Research}
\subsection{Inverted Index}

An inverted index is a data structure used for text retrieval, which maps keywords to the corresponding document identifiers and their positions. It allows for direct access to the document identifiers and positions where the indexed keywords occur. The inverted index consists of four components: terms (indexed keywords), document identifiers ($doc$s), term frequency ($TF$), and positions of the indexed keywords within the documents ($loc$s). These components together form an inverted index entry, which is the fundamental unit of the inverted index structure. In practical retrieval scenarios, the ability to locate documents or records based on given keywords or attributes is crucial, and the inverted index enables fast retrieval of documents and content containing specific keywords.

\begin{figure}[H]
  \centering
  \includegraphics[width=0.5\textwidth]{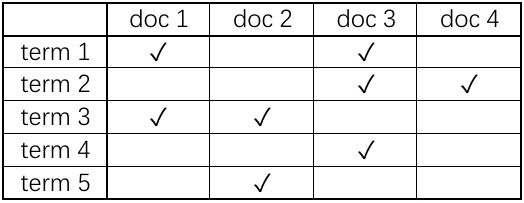}{}
  \caption{Binary Structure of "Keyword-Document" Relationship}
  \label{fig:fig1}
\end{figure}

Specifically, the inverted index data structure comprises two main parts: the lexicon and the inverted lists. Each inverted posting in the inverted list contains additional information, such as the term frequency \cite{a5}. Furthermore, an inverted index entry, representing a keyword and its corresponding inverted postings, can be expressed as: 

$$
term_i \rightarrow \{doc_{(i,1)},TF_{(i,1)},(loc_1,loc_2,\dots)\},\{doc_{(i,2)},TF_{(i,2)},(loc_1,loc_2,\dots)\},\dots
$$

The lexicon records all unique words in the text, while the inverted lists record the occurrence positions of each word in the text. This data structure significantly improves the speed of text retrieval and meets the real-time response requirements for user queries involving aggregation and ranking. The lexicon is a crucial component of the inverted index, maintaining relevant information for all keywords in the document collection. During the index construction process, each keyword is assigned a unique identifier to represent it. Each entry in the lexicon contains keyword information, logical pointers to the corresponding inverted lists, and other relevant information. The inverted lists maintain the association between documents and keywords, consisting of identifiers for all documents containing the keyword. During the index construction process, each document is assigned a unique identifier to represent it.

\begin{figure}[H]
  \centering
  \includegraphics[width=0.7\textwidth]{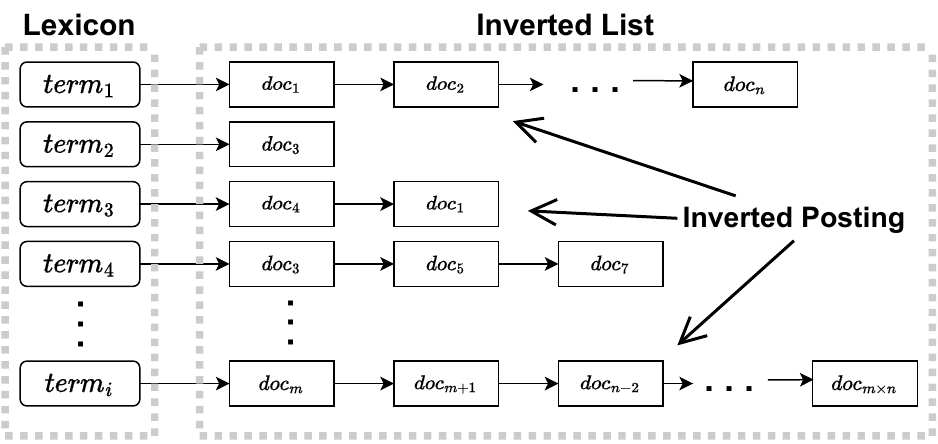}{}
  \caption{Illustration of Inverted Index Structure}
  \label{fig:fig2}
\end{figure}

\subsection{Co-occurrence Networks}
Co-occurrence networks, as a text analysis method, effectively uncover the connections between vocabulary and the structure of scientific knowledge. By constructing co-occurrence networks, we can gain a better understanding of the relationships between different topics and the evolution of knowledge systems.

\begin{figure}[H]
  \centering
  \includegraphics[width=1\textwidth]{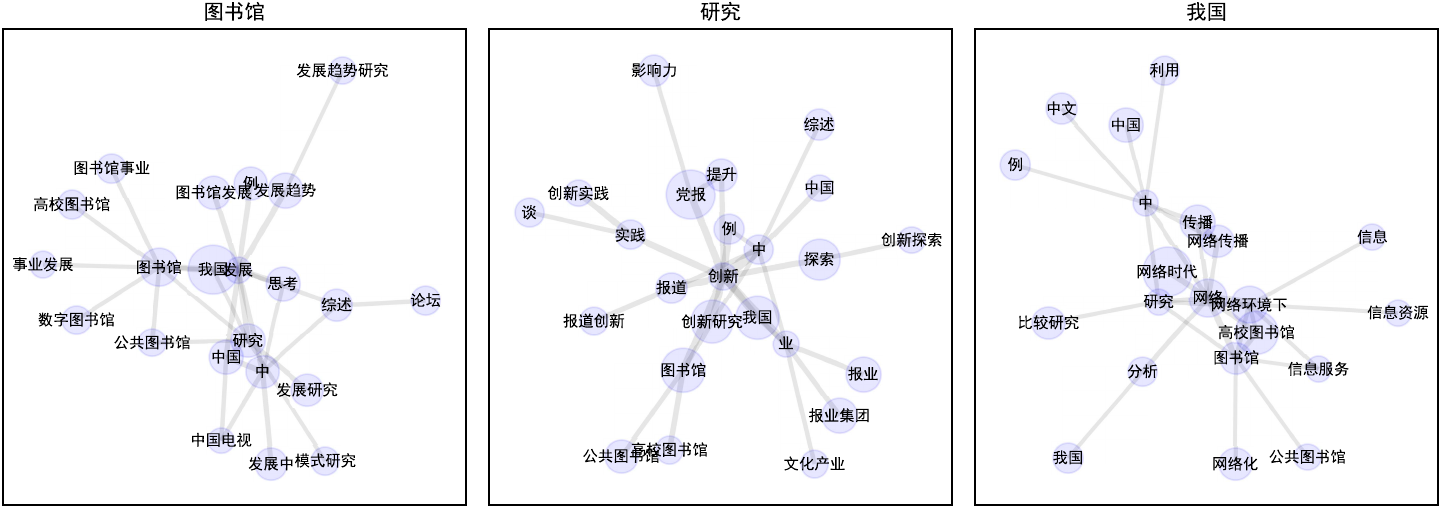}{}
  \caption{Illustration of Co-occurrence Networks under Different Topics}
  \label{fig:fig3}
\end{figure}

Specifically, a co-occurrence network is an undirected weighted network where the strength of relationships between vocabulary terms is not uniform. The more frequently two vocabulary terms co-occur in different articles, the stronger their relationship. During the construction of co-occurrence networks, it is assumed that the weight of co-occurrence between any two terms is the same. In other words, whether terms A and B co-occur in article A or article B, their co-occurrence contributes the same weight to the relationship between A and B. This construction method implies that the focus of co-occurrence network research is not on the value of specific term pairs in individual articles but on the overall structure of the co-occurrence network \cite{a4}.

Co-occurrence networks, also known as keyword co-occurrence networks, typically refer to networks formed by keywords and their co-occurrence relationships in articles. In such networks, nodes represent keywords, and edges indicate the simultaneous appearance of two words in a document. Keyword co-occurrence can be considered as representing a relevant topic or belonging to a cluster of related topics. Such networks can reveal the entity relationship characteristics within the scientific knowledge system and the growth patterns of scientific knowledge concepts.

\subsection{Co-occurrence Network Construction Algorithms}
In this section, we will analyze in detail the mainstream traversal-based algorithms for constructing co-occurrence networks \cite{a4, a6, a7}. Traversal-based construction algorithms involve the following steps: extracting relevant documents and performing tokenization, traversing each document and iterating over each pair of vocabulary terms, and constructing a sparse matrix based on the co-occurrence counts of term pairs in the documents. The algorithm can be described using pseudocode as Algorithm 1:

Assuming there are $n$ documents and an average of $m$ terms per document, the time complexity of the traversal algorithm is $O(nm^2)$ since we need to traverse each term in each document and its co-occurring terms. In terms of space complexity, as we need to store the co-occurrence relationships between each term and other terms, the sparse matrix will occupy a fixed space of $O(m^2)$.

\begin{algorithm}[H]
\caption{Traversal-based Co-occurrence Network Construction}
\SetAlgoLined
\KwIn{Filtering conditions}
\KwOut{Co-occurrence matrix}
Document collection $\gets$ Documents in the database that satisfy the filtering conditions\;
Initialize co-occurrence matrix\;
\ForEach{document in the document collection}{
    Tokenized result $\gets$ Tokenize the document using a tokenizer\;
    Term pairs $\gets$ Generate pairs of terms from the tokenized result\;
    \ForEach{term pair in the term pairs}{
        \If{term0 and term1 in the term pair are the same}{
            Skip\;
        }
        Increment co-occurrence matrix[term0, term1]\;
    }
}
\Return co-occurrence matrix\;
\end{algorithm}

In practical applications, traversal-based construction methods suffer from high time consumption due to the need for tokenization during execution, and the sparse matrix consumes a significant amount of memory space. Additionally, the obtained co-occurrence network is often too comprehensive, making it susceptible to interference from low-frequency terms and resulting in weak robustness of the final results.

\section{Algorithm for Co-occurrence Network Construction Based on Inverted Index}
This algorithm incorporates the ideas of inverted index and breadth-first search into the construction of co-occurrence networks. By utilizing the inverted index, the tokenization process is separated from the traditional methods, significantly reducing the time complexity constant. Moreover, it allows for quick identification of documents based on a given term, facilitating the identification of co-occurring terms.

\begin{figure}[H]
  \centering
  \includegraphics[width=0.6\textwidth]{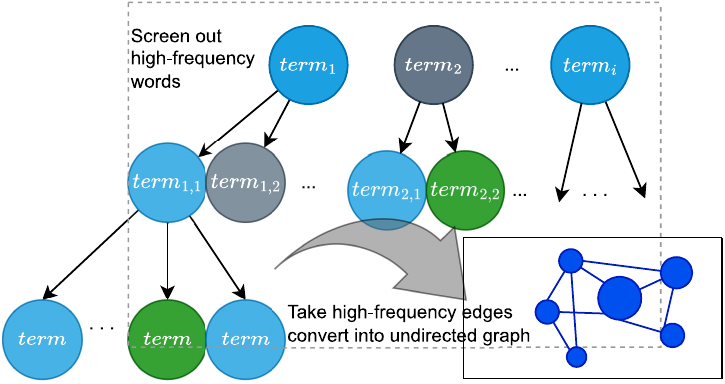}{}
  \caption{Illustrates the optimized algorithm}
  \label{fig:fig4}
\end{figure}

The algorithm can be roughly divided into three steps. Firstly, it retrieves the words and their frequencies from the documents that meet the filtering conditions using the inverted index. Then, it iterates through the retrieved words, adding each word to the retrieval conditions one by one. This process generates new high-frequency words. The algorithm stores the retrieved word and its corresponding high-frequency word as an edge pair until the depth limit is reached.

The algorithm has several implementation methods. This paper presents two implementations: one based on recursive depth-first search and the other based on breadth-first search. In this section, we will introduce these two implementation approaches separately and demonstrate that the algorithm is not sensitive to the newly introduced parameter, "search depth."

\subsection{Recursive Implementation Method}
One relatively simple implementation of the algorithm is to use a recursive approach, which includes the following steps: obtaining the inverted index of relevant documents based on the query conditions, retrieving the most frequent words from the inverted index, and calling the algorithm again by adding each word to the retrieval conditions.

Assuming that there are n high-frequency words that satisfy the search conditions in each retrieval, and the search depth is d, the time complexity of the recursive construction method based on the inverted index is $O(n^d)$. However, it is important to note that there is a process of extracting high-frequency words, which requires traversing the word set. Therefore, the actual time complexity is $O(n^2d)$. This time complexity is not necessarily lower than that of a traversal-based algorithm (Algorithm 2).

\begin{algorithm}[H]
\SetAlgoLined
\caption{Recursive Construction of Co-occurrence Network}
\KwIn{Filtering conditions, search depth}
\KwOut{}{Node and edge records}
Initialize node and edge records\;
\If{recursion depth is reached}{
    \Return node and edge records\;
}
Word set $\gets$ Words from the inverted index that satisfy the filtering conditions\;
High-frequency word set $\gets$ Extract high-frequency words from the word set\;
\ForEach{word in the high-frequency word set}{
    New filtering conditions $\gets$ Add the word to the filtering conditions\;
    Add the word from the filtering conditions and the word to the node and edge records\;
    Recursive Construction of Co-occurrence Network using the new filtering conditions\;
}
\end{algorithm}

Compared to a direct traversal approach, the time consumption of this method depends to some extent on the search depth, $d$. In terms of space complexity, each recursive call requires allocating new space for the stack, resulting in a space complexity of $O(d)$. However, in most programming languages, especially in scripting languages, recursive calls consume a significant amount of storage resources. Therefore, although this algorithm essentially only stores the required results, its recursive nature leads to higher space and time complexity constants in practical applications.

It is important to note that the recursive search should not start immediately after obtaining the word set. Instead, it is necessary to retrieve the high-frequency words based on word frequency. Using high-frequency word search has the following benefits: it reduces the search for low-frequency words, resulting in a co-occurrence network that better reflects the document's theme and semantic information. Additionally, it improves performance, as increasing the number of searches would lead to a growth in the $n$ term in the time complexity, resulting in excessive performance overhead.

\subsection{Algorithm Insensitivity to Search Depth}
Based on the analysis above, it is evident that the new algorithm has a higher complexity compared to traditional traversal algorithms in terms of its structure. However, what causes the new algorithm to be faster in practice? By examining the time complexity of the new algorithm, we can observe that its more complex part mainly lies in the parameter $d$, which represents the search depth. The search depth can lead to an exponential increase in time complexity.

\begin{figure}[t]
  \centering
  \includegraphics[width=1\textwidth]{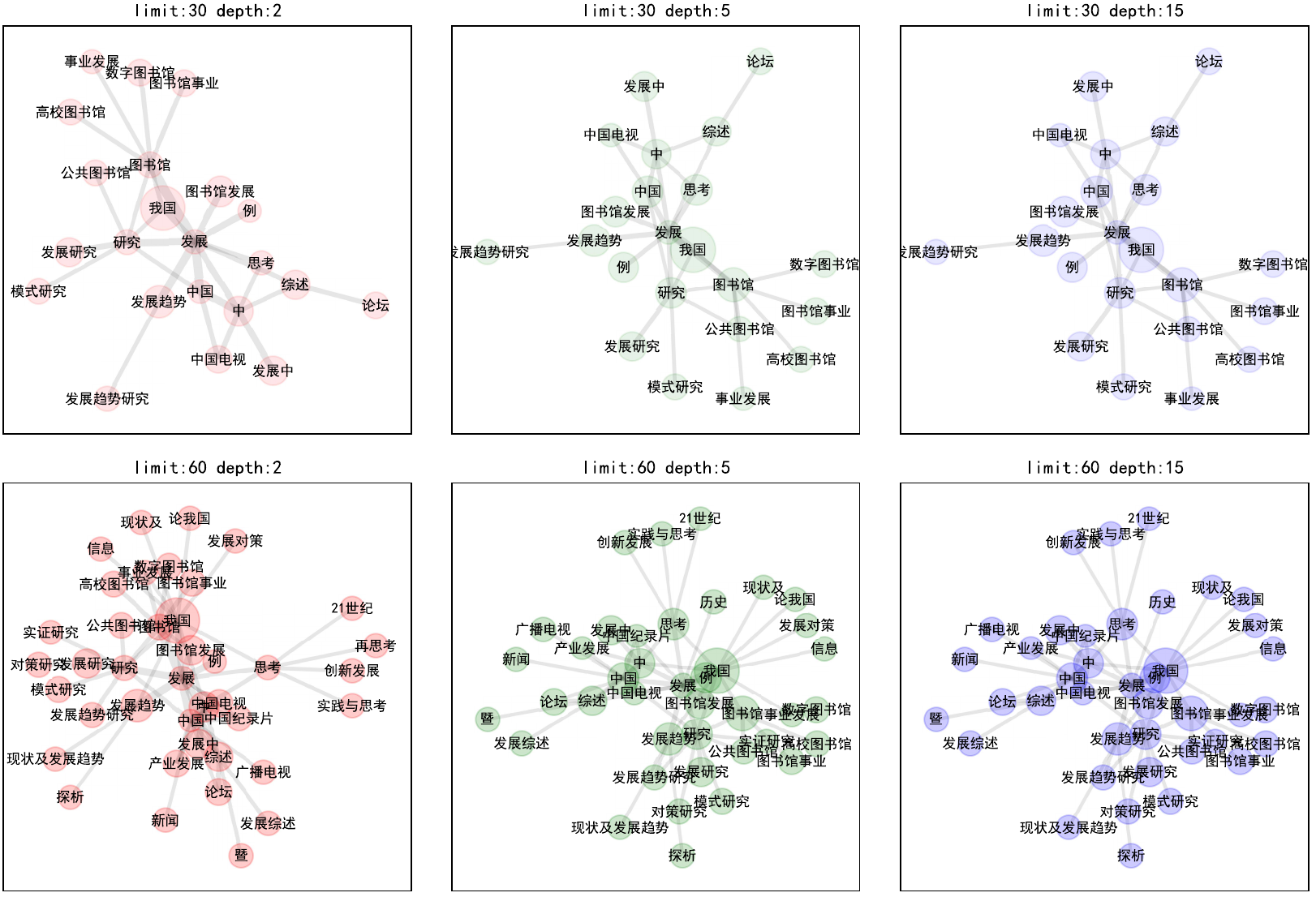}{}
  \caption{Visualization Results Comparison for Different Edge Quantity Restrictions and Search Depths}
  \label{fig:fig5}
\end{figure}

As shown in Figure \ref{fig:fig5}, the visualization results vary with different search depths and edge quantity restrictions. The "limit" refers to the maximum number of edges plotted during visualization, while "depth" represents the search depth. When the limit is set to 60, the information in the graph is rich, but it becomes difficult for the naked eye to discern the data, resulting in less than ideal visualization. Comparing the co-occurrence network results for depth 2 and depth 5, we can observe some differences, with the depth 5 network displaying richer connections. However, when comparing the networks for depth 5 and depth 15, despite the threefold increase in depth, the generated network results do not differ significantly.

The above example partially illustrates that the restriction on the number of edges during visualization can affect the construction of the network. In shallow search depths, the network structure can be influenced. However, once the depth reaches a certain level, it does not significantly impact the algorithm's results. In other words, the algorithm is insensitive to the search depth. Based on the complexity analysis mentioned earlier, a large search depth would result in a geometric increase in time consumption. Therefore, the search depth should not be a variable but rather the minimum value that ensures network completeness.

\subsection{Improvement Using Breadth-First Search}
The analysis of this algorithm is essentially consistent with the recursive algorithm. This implementation also requires specifying the search depth, which to some extent determines the scale of the network. The pseudocode for this implementation is described as follows:

\begin{algorithm}[H]
\SetAlgoLined
\caption{Co-occurrence Network Construction Algorithm Based on Inverted Index}
\KwIn{Filtering conditions, search depth}
\KwOut{Node and edge records}
Initialize node and edge records\;
\While{not reached search depth}{
    Word set $\leftarrow$ Words from the inverted index that satisfy the filtering conditions\;
    High-frequency word set $\leftarrow$ Extract high-frequency words from the word set\;
    \ForEach{word in the high-frequency word set}{
        New filtering conditions $\leftarrow$ Add the word to the filtering conditions\;
        Add the word from the filtering conditions and the word to the node and edge records\;
    }
}
\Return{Node and edge records}
\end{algorithm}

Assuming that there are n high-frequency words that meet the search criteria in each retrieval, and the search depth is d, as mentioned earlier, the algorithm complexity is $O(n^2d)$. However, based on the sensitivity analysis of the search depth, it can be treated as a constant. Therefore, the actual algorithm complexity is $O(n^2)$, which is a significant improvement compared to the traditional algorithm's $O(nm^2)$.

Compared to the recursive implementation method, breadth-first search directly loops based on the search depth, avoiding the performance overhead of recursive function calls. Additionally, the loop structure, rather than recursion, also allows for better utilization of computer hardware optimization based on the principle of program locality. It also facilitates further optimization of I/O performance through the use of multi-threading or coroutine programming techniques.

Compared to the recursive approach, breadth-first search offers several advantages. Firstly, it eliminates the overhead of recursive function calls, resulting in improved performance. Secondly, the iterative nature of breadth-first search allows for better utilization of hardware resources and optimization techniques. This includes leveraging the principles of program locality, which can enhance cache efficiency and reduce memory access latency. Additionally, the iterative structure of breadth-first search lends itself well to parallelization, enabling the use of multiple threads or coroutines to further enhance performance.

In summary, the breadth-first search approach provides an improvement over the recursive implementation. By directly looping based on the search depth, it avoids the performance overhead associated with recursion and allows for better utilization of hardware resources. These advantages contribute to improved algorithm complexity and the potential for further optimization in terms of I/O performance and parallelization.

\section{Experiment}
To validate the actual performance improvement brought by the algorithm, experiments were designed in this study. The experiments involved extracting high-frequency words from the dataset and using these words as filtering conditions to retrieve documents. The traditional traversal algorithm and the optimized algorithm proposed in this paper were then used to construct co-occurrence networks. Performance metrics such as runtime and memory usage were recorded and analyzed for each algorithm.

\subsection{Dataset}
\begin{figure}[H]
  \centering
  \includegraphics[width=1\textwidth]{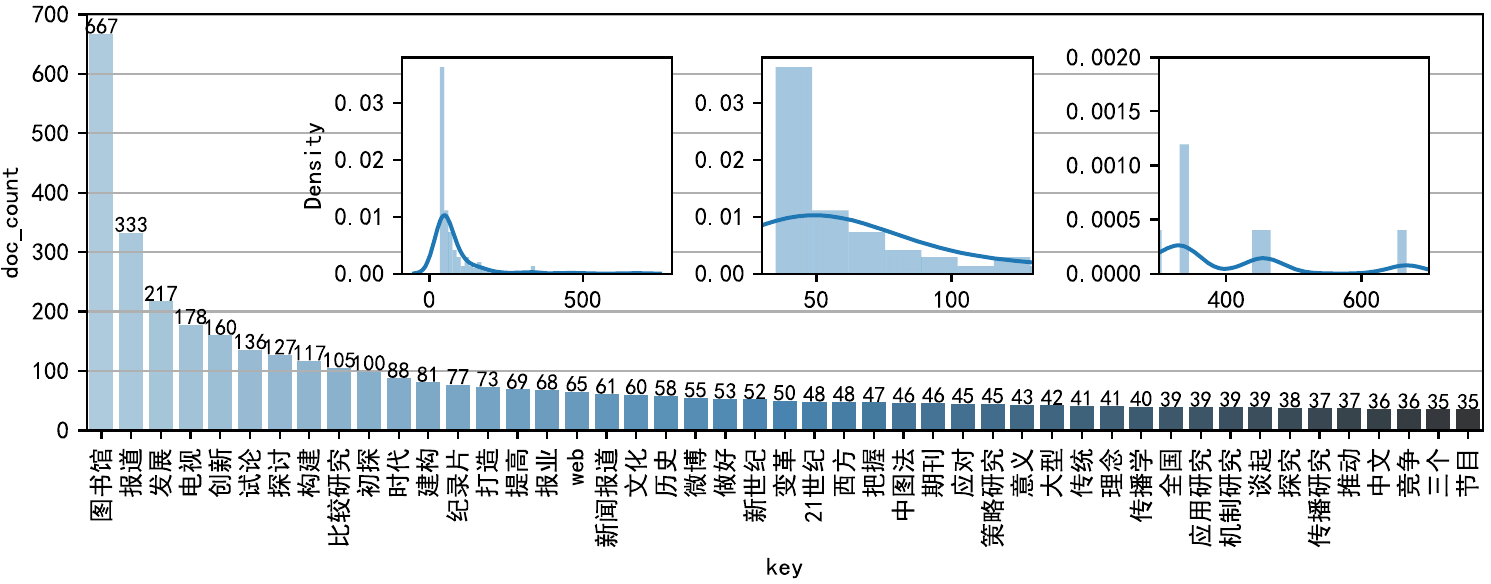}{}
  \caption{Data Distribution}
  \label{fig:fig6}
\end{figure}

The experiment utilized the CSL (Chinese Scientific Literature) dataset, a large-scale Chinese scientific literature dataset obtained from the "Qianyan" open-source NLP platform \cite{a8}. The CSL dataset consists of 396,209 Chinese core journal papers' metadata, including titles, abstracts, keywords, disciplines, and categories. The CSL dataset covers journal papers published from 2010 to 2020, selected based on the Chinese Core Journal Catalog, and annotated with discipline and category labels. It is divided into 13 categories (primary labels) and 67 disciplines (secondary labels).

Due to time and computational constraints, this study used a subset of the dataset with discipline categories "Library, Information, and Archive Management" and "Journalism and Communication" as the specific experimental data. Firstly, the keywords column of the entire dataset was extracted to obtain the keyword information of all the papers. The extracted keywords were then used as domain expansion dictionaries for segmentation, while the stopword dictionaries from Harbin Institute of Technology, Baidu, and Sichuan University were used as stopwords\cite{a9,a10}. Finally, the title and content of the dataset were indexed using Elasticsearch, and IK Analyzer was employed for segmentation.

The segmented and indexed data were aggregated for statistical analysis, and the distribution was visualized as shown in the figure \ref{fig:fig6}. The large graph displays a partial list of words sorted from small to large based on their frequency. The horizontal axis represents specific words, while the vertical axis represents the count of documents in which they appear. It can be observed that high-frequency words are not the most widely distributed, as most words are distributed below 50. The three smaller graphs in the middle depict the distribution based on the number of words, with the horizontal axis representing the number of words and the vertical axis representing the probability of their distribution. Overall, the distribution is mainly concentrated below 50 words. The two enlarged graphs on the right show the low-frequency and high-frequency parts separately, revealing the presence of a certain number of high-frequency words. In general, the distribution of words approximately follows a Poisson distribution, which is reasonable and suitable for experimental data.

\subsection{Environment and Parameters}
In terms of hardware, the experiments were conducted on a personal portable laptop with an AMD Ryzen 7 4800H processor, with a base speed of 2.9GHz. The laptop had 2×8GB 3200MHz memory and a solid-state drive (SSD) with the model SAMSUNG MZVLB512HBJQ-000L2.

Regarding software, IK Analyzer and Elasticsearch were used for segmentation and indexing, both at version 8.6.1. The traditional algorithm and the optimized algorithm were implemented using Python, with the interpreter version being 3.9.13 on the Windows platform. Apart from the Elasticsearch client, no other external libraries were used in the algorithm implementation. The statistical analysis of the experimental results was performed using the Pandas\cite{a11}, SciPy\cite{a12}, and Seaborn\cite{a13} libraries, with versions 2.0.0, 1.10.1, and 0.12.2, respectively. The implementation of the algorithms and the experimental code have been made open-source and can be obtained from \href{https://gitee.com/jaydencheng/cooccurrence-inverted-index}{this page}\footnote[1]{\href{https://gitee.com/jaydencheng/cooccurrence-inverted-index}{https://gitee.com/jaydencheng/cooccurrence-inverted-index}}.

\subsection{Experimental Results Analysis}
\begin{figure}[H]
  \centering
  \includegraphics[width=0.9\textwidth]{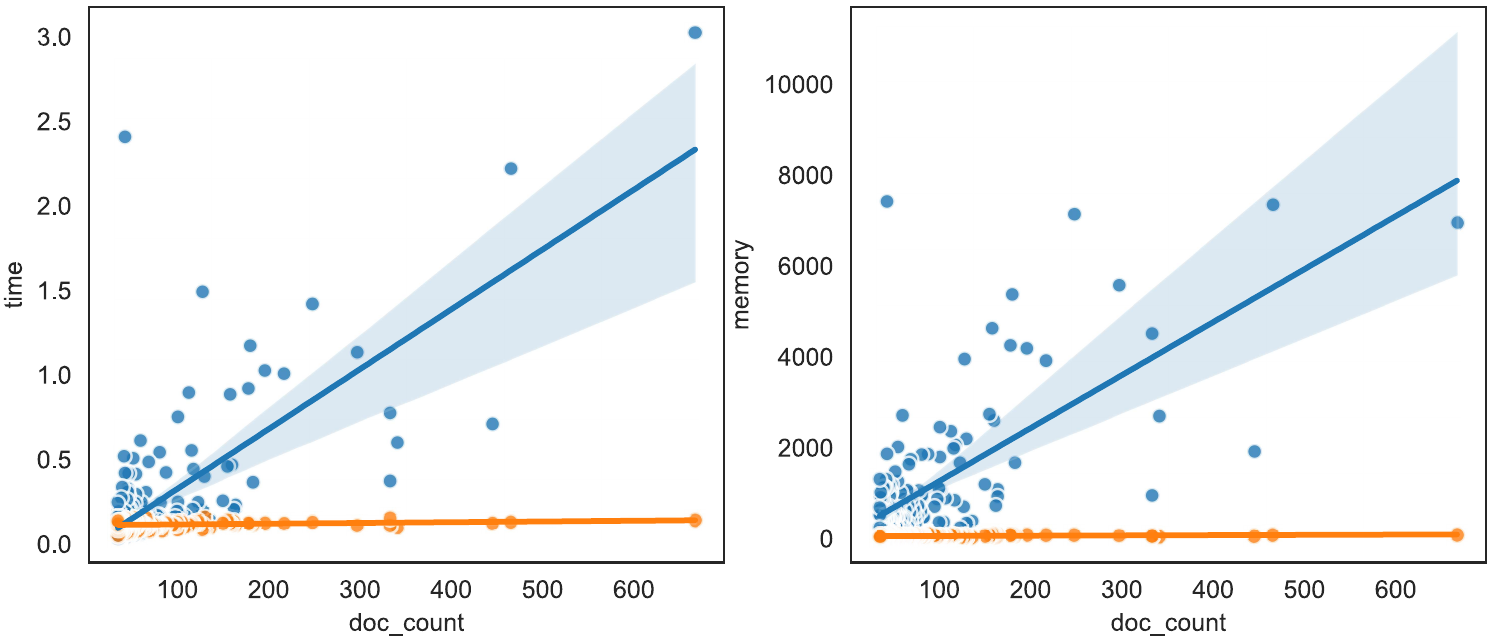}{}
  \caption{Scatter Plot of Experimental Results}
  \label{fig:fig7}
\end{figure}

The experimental results are presented in Figure \ref{fig:fig7}, where algorithms' performance is depicted with confidence intervals. The algorithms' performance is shown in terms of time and memory consumption.

The left graph represents the time taken by the algorithms to construct the co-occurrence network, measured in seconds (s). The right graph represents the memory space consumed by the algorithms, measured in bytes (byte). The experimental results clearly demonstrate that the proposed algorithm not only consumes less time and space but also exhibits greater stability.

\begin{figure}[H]
  \centering
  \includegraphics[width=0.9\textwidth]{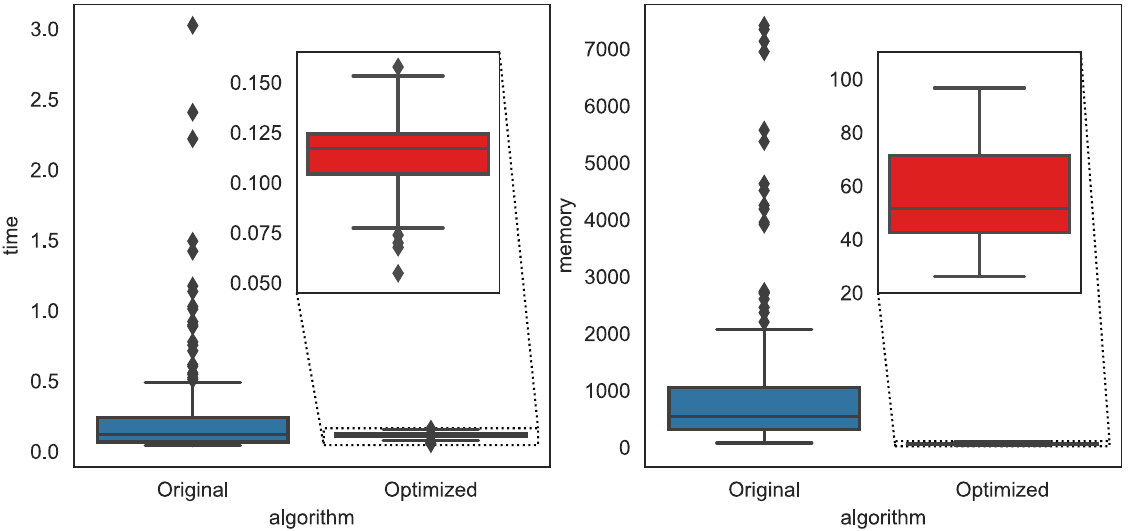}{}
  \caption{Box Plot of Experimental Results}
  \label{fig:fig8}
\end{figure}

Statistical methods can provide further insights. Box plots are used to visualize the results of both algorithms. It is evident that the traditional algorithm has many outliers, indicating the stability of the new algorithm. The new algorithm has lower average values in terms of time and memory consumption compared to the traditional algorithm. Particularly in terms of time, the new algorithm consistently performs below 0.16 seconds in the experimental environment, meeting the requirements of web systems.

Since the experimental data does not follow a normal distribution, non-parametric tests are used to examine whether there are significant differences between the performance before and after optimization. The null hypothesis (H0) is that the performance data distributions are the same before and after optimization, while the alternative hypothesis (H1) is that they are different. The Wilcoxon and Mann-Whitney tests are conducted, yielding test statistics and p-values of 5809.0, 5.8e-5, and 34930.0, 1.6e-62, respectively. As all p-values are less than 0.001, the null hypothesis is rejected, indicating that the performance of the new algorithm is significantly better than that of the traditional traversal algorithm.

\section{Conclusion}
This study proposed and implemented an optimized algorithm for constructing co-occurrence networks based on inverted indexing and breadth-first search. This algorithm requires an additional parameter, the search depth, which is used to obtain a high-frequency word set based on the depth and inverted index. The algorithm then continues to expand the word set by adding words to the filtering conditions, ultimately constructing the nodes and edges of the co-occurrence network by combining pairs of words from the word set. Theoretical analysis and experimental results demonstrate that this algorithm outperforms traditional approaches in terms of both time and space complexity.

Future research can explore whether other text data storage structures can further enhance the construction of co-occurrence networks. Additionally, investigating the impact of the "search depth" parameter and the dataset on the performance of co-occurrence network construction would be valuable. Furthermore, parallelizing the algorithm could be explored to achieve even greater performance improvements.

\section*{Acknowledgments}
This work is a part of the "Big Data Intelligent Topic Selection Platform for Research Projects". I would like to express my sincere gratitude to Professor Changbin Jiang for his guidance and support throughout this research endeavor. His expertise, mentorship, and encouragement have been invaluable in the successful completion of this paper.

I would also like to extend my appreciation to Professor Xuefeng Wang for his guidance and support during the course of "Information Organization". 

\bibliographystyle{ieeetr}  
\bibliography{references}

\begin{thebibliography}{10}

\bibitem{a1}
H.~Zewen, S.~Jianjun, and W.~Yishan, ``Research review on application of
  knowledge mapping in china,'' {\em Libary and Information Service}, vol.~57,
  no.~03, p.~131, 2013.

\bibitem{a2}
G.~Yunjun, G.~Congcong, G.~Yuxiang, and C.~Lu, ``Survey on data integration
  technologies for relational data and knowledge graph,'' {\em Journal of
  Software}, vol.~34, no.~5, p.~2365, 2023.

\bibitem{a3}
Z.~Xiaofei, C.~Hangxing, and Z.~Chunhua, ``Research on micro-blog subject word
  extraction based on semantic concept and word co-occurrence,'' {\em
  Information Science}, vol.~39, no.~1, pp.~142--147, 2021.

\bibitem{a4}
Z.~Shuan, W.~Xi, D.~Jipeng, S.~Yi, and S.~Rencheng, ``Subject words extraction
  algorithm based on keyword co-occurrence network,'' {\em Complex Systems and
  Complexity Science}, vol.~20, no.~1, pp.~74--80, 2023.

\bibitem{a5}
W.~Lai, Z.~Tengda, W.~Zhengfei, and H.~Jiaming, ``Construction of inverted
  index for encrypted electronic medical records search based on bloom filter
  and b+ tree,'' {\em Computer Applications and Software}, 2021.

\bibitem{a6}
W.~Jinming, H.~Yuefang, and C.~Lei, ``Automatic expression of co-occurrence
  clustering based on indexing rules of medical subject headings,'' {\em Data
  Analysis and Knowledge Discovery}, vol.~4, no.~9, pp.~133--144, 2020.

\bibitem{a7}
H.~Zhiqiang, W.~Lipeng, and Z.~Pengyun, ``Research and improvement of keyword
  extraction algorithm based on word co-occurrence,'' {\em Electronic
  Technology and Software Engineering}, no.~1, pp.~144--146, 2018.

\bibitem{a8}
Y.~Li, Y.~Zhang, and et~al., ``{CSL}: A large-scale {C}hinese scientific
  literature dataset,'' in {\em Proceedings of the 29th International
  Conference on Computational Linguistics}, (Gyeongju, Republic of Korea),
  pp.~3917--3923, International Committee on Computational Linguistics, Oct.
  2022.

\bibitem{a9}
G.~Qin, D.~Sanhong, and W.~Hao, ``Chinese stopwords for text clustering: A
  comparative study,'' {\em Data Analysis and Knowledge Discovery}, vol.~1,
  no.~3, pp.~72--80, 2017.

\bibitem{a10}
Y.~Juan, Y.~Jidong, and F.~Shu, ``Research on synonym recognition method based
  on syntactic structure analysis,'' {\em Journal of Modern Information},
  no.~9, pp.~35--40, 2013.

\bibitem{a11}
{W}es {M}c{K}inney, ``{D}ata {S}tructures for {S}tatistical {C}omputing in
  {P}ython,'' in {\em {P}roceedings of the 9th {P}ython in {S}cience
  {C}onference} ({S}t\'efan van~der {W}alt and {J}arrod {M}illman, eds.),
  pp.~56 -- 61, 2010.

\bibitem{a12}
P.~Virtanen, R.~Gommers, and et~al., ``{{SciPy} 1.0: Fundamental Algorithms for
  Scientific Computing in Python},'' {\em Nature Methods}, vol.~17,
  pp.~261--272, 2020.

\bibitem{a13}
M.~L. Waskom, ``seaborn: statistical data visualization,'' {\em Journal of Open
  Source Software}, vol.~6, no.~60, p.~3021, 2021.

\end{thebibliography}

\end{document}